\documentclass[fleqn,10pt]{wlscirep}
\usepackage[utf8]{inputenc}
\usepackage[T1]{fontenc}
\usepackage{algorithm}
\usepackage[noend]{algpseudocode}

\usepackage{graphicx}

\usepackage{epstopdf}
\DeclareGraphicsExtensions{.pdf,.png}

\usepackage{subfig}
\usepackage{multicol}
\usepackage{xcolor}

\usepackage{booktabs}
\usepackage{ upgreek }

\usepackage{tikz}
\usepackage{xcolor}

\tikzstyle{line} = [draw, -latex']

\usetikzlibrary{positioning,chains,fit,shapes,calc,arrows, decorations, bayesnet}

\title{Network characteristics of financial networks}

\author[1,2,*]{M. Boersma}
\author[1]{S. Sourabh}
\author[2]{L.A. Hoogduin}
\author[1]{D. Kandhai}
\affil[1]{Computational Science Lab, University of Amsterdam, Amsterdam, the
Netherlands}
\affil[2]{KPMG, Amstelveen, the Netherlands}

\affil[*]{m.boersma@uva.nl}

\begin{abstract}
We embrace a fresh perspective to auditing by analyzing a large set of companies as complex financial networks rather than static aggregates of balance sheet data. Preliminary analyses show that network centrality measures within these networks could significantly enhance auditors' insights into financial structures. Utilizing data from over 300 diverse companies, we examine the structure of financial statement networks through bipartite graph analysis, exploring their scale-freeness by comparing degree distributions to power-law and exponential models. Our findings indicate heavy-tailed degree distribution for financial account nodes, networks that grow with the same diameter, and the presence of influential hubs.  This study lays the groundwork for future auditing methodologies where baseline network statistics could serve as indicators for anomaly detection, marking a substantial advancement in audit research and network science.

\end{abstract}
\begin{document}

\flushbottom
\maketitle

\thispagestyle{empty}

\section*{Introduction}
Auditing is a critical service where a third party evaluates the truthfulness of a company's financial information -- a market valued at 217 billion~\cite{business_wire_2020}. Despite the importance of trust in financial reporting like balance sheets, current audit practices remain predominantly manual, with algorithmic procedures scarcely integrated. The availability of detailed transaction data offers the opportunity to analyze financial data differently. In other words, balance sheets present a static, year-end snapshot of a company's financial stance, failing to encapsulate the dynamic flow of funds through its accounts. A novel approach, as Boersma et al. (2018)~\cite{Boersma2018} proposes, envisions companies as intricate financial networks rather than mere collections of balance sheet figures. This network-centric view, characterized as a bipartite graph, captures not only the financial position but also the transactional architecture and the circulation of monetary resources driven by business operations.

\textcolor{black}{Applying network analysis to these structures presents a significant potential for practical applications in auditing. A network representation increases the auditor's understanding of the complexity of the financial structure within a company~\cite{Boersma2018}. Additionally, network measures ~\cite{Freeman1978-ce, girvan2002community, Albert2002-hm} would allow an auditor to focus on critical parts within the network.
And finally, comparing network measures of companies within a sector will potentially enable an auditor to  identify outliers and their potential ramifications on the risk assessment.} In general, this raises the question if we could use network \textcolor{black}{analysis} to understand whether, for example, a company is in a healthy shape? Or identify network parts with higher audit risk? The insights could ultimately lead to new data-driven audit approaches. Ambitious, but not much different from a chemist or biologist analyzing a chemical network structure to infer a molecule's properties~\cite{Bonneau2008-iv,Gilmer2017-il} \textcolor{black}{or analyzing the structure in biological-, brain-networks, or a sociologist analyzing social-networks}~\cite{girvan2002community, Albert2002-hm, bullmore2009complex}. However, before we pursue such ambitious ideas, we must understand the basic network properties. 

Numerous network statistics have been proposed in the literature ~\cite{Freeman1978-ce, girvan2002community, Albert2002-hm, barabasi1999emergence, Newman2003-hc}. The seminal work of \cite{watts1998collective} introduced a set of interesting statistics that are observed in most real networks. Here we analyze a sizable set of financial statements networks and focus on these network metrics, i.e., betweenness centrality, closeness centrality, degree centrality, and degree distribution. In our study, however, we use a special type of network, a bipartite network. Complex systems that have a bipartite network structure are plentiful, for example, in economics: in companies sharing board members~\cite{Robins2004,Conyon2004-jm,Battiston2004-ht,Newman2001-ik,Seierstad2011-bg}, in management science~\cite{Kogut2007-pl}, and in financial networks~\cite{Caldarelli2004-cr,Dahui2006-bk, Garlaschelli2005-ml}. As well as in the social domain where people, for example, go to different locations~\cite{Eagle2006-yj,Faust2002-ie}, or rate products~\cite{Ziegler2005-wl}, or collaborate on scientific work ~\cite{newman2001scientificb,Roth2005-ui}. All are examples of bipartite networks because the network has two modes. 

The statistics mentioned above might not work on a bipartite network, as argued in Refs. ~\cite{Borgatti1997, Latapy2008-ty, Borgatti2009-hr}. 
There are two ways to analyze a bipartite network: (1) derive a one-mode projection from the bipartite network and analyze this network, (2) define new statistics that respect the bipartite nature of the network. A one-mode projection results in (i) information loss, (ii) or results that are an artifact of the projection. Therefore, we use the network measures devised by Borgatti et al.~\cite{Borgatti1997} to avoid the aforementioned drawbacks. These measures consider the bipartite nature of the network and adjust the definition of network measures accordingly, as detailed in Section~\ref{sec:centrality}. 

The focus of the research is the network statistics for a large set of financial statements networks. More specifically we:
\begin{enumerate}
    \item use real transaction data of 300+ companies from different sectors to analyze the financial statements network structure
    \item provide an overview of bipartite network statistics and discuss these for a large set of financial statements networks
    \item we investigate whether financial statements networks are scale-free by fitting their degree distributions to  power-law and exponential distributions
\end{enumerate}
With this, we provide a valuable contribution to the network science and audit research community. First, we explain the financial statements network dataset in Section~\ref{sec:fin:dataset}. Next, we study basic network statistics (the size of the network and the centrality measures) in Section~\ref{sec:basicnetworkstats}.  In Section~\ref{sec:degreedistribution}, we investigate the degree distributions. And finally, in Section~\ref{sec:conclusion}, we summarize our main findings.

\section{The financial statements networks dataset}
\label{sec:fin:dataset}
\textcolor{black}{In this section, we briefly introduce the financial statements network dataset. The dataset consists of 300+ journal entry datasets of companies, each dataset ranges from a single journal entry up to 5 million journal entries per company. The companies are mostly located across Europe and are active in 28 industries. Table~\ref{table:industrycoverage} shows the top 5 industries -- other industry categories contain between 1 and 12 companies.} 
Two important remarks related to the dataset impact the results: (i) it is classified as highly confidential, (ii) it is collected during live audit engagements. Consequently, because we obtained a snapshot of the live data, we expect some noise to be present in the dataset -- for example, the small journal entry datasets. Nonetheless, we believe that we can obtain relevant statistics from this dataset due to the significant size. The second part, confidentiality, implies that we may only analyze the dataset on secure hardware.  Due to the limited computational power of the restricted hardware, we limited our analysis to networks with a maximum of 100.000 nodes. As a result, we analyzed 285 companies instead of the full 312 we obtained.  

\begin{figure}
    \centering
    \includegraphics[width=0.40\textwidth]{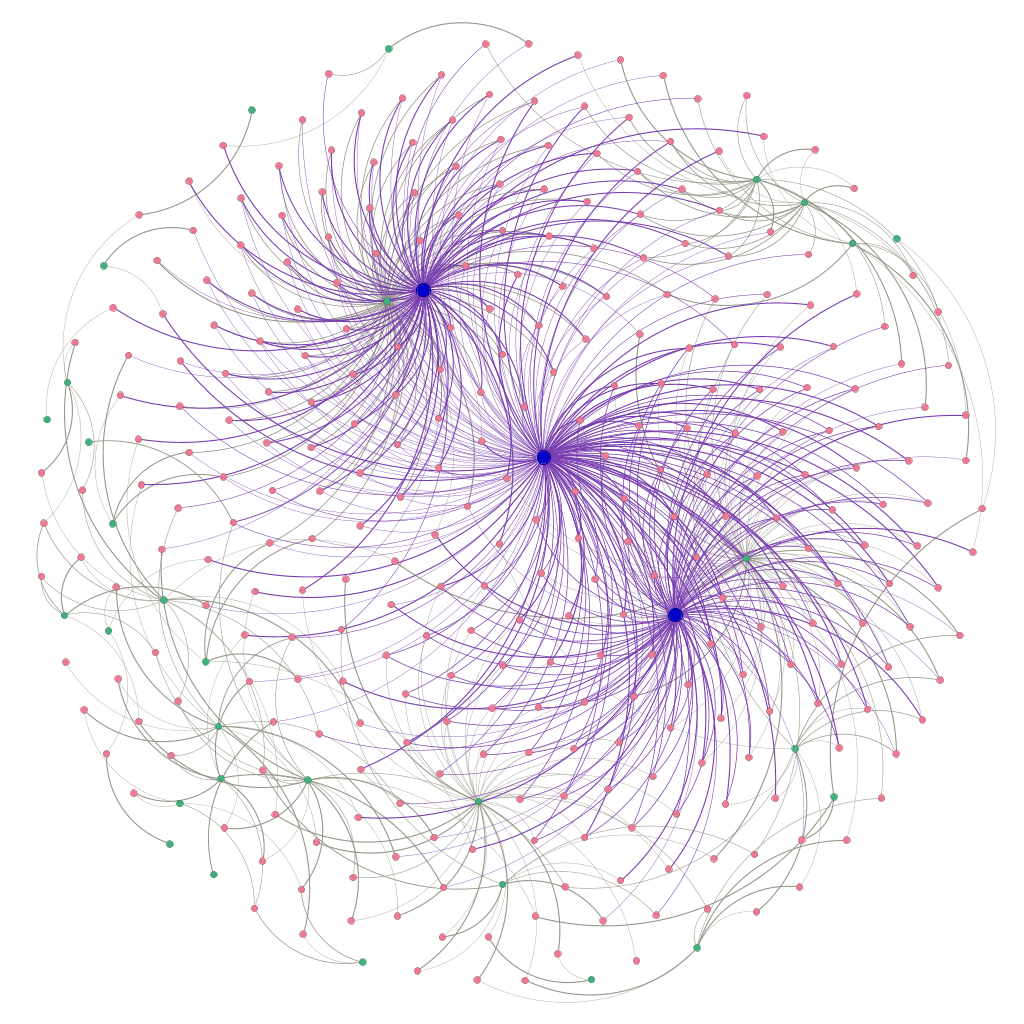}
    \includegraphics[width=0.40\textwidth]{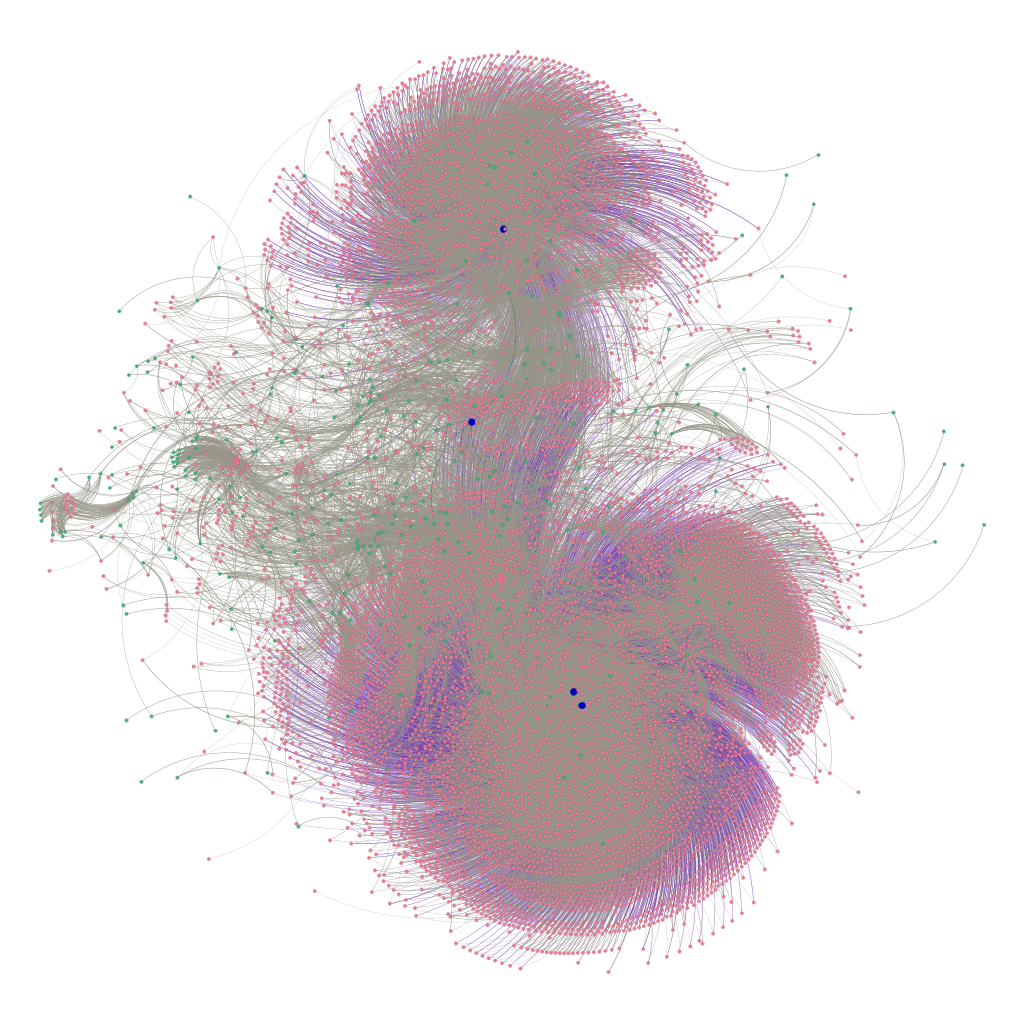}
    \includegraphics[width=0.40\textwidth]{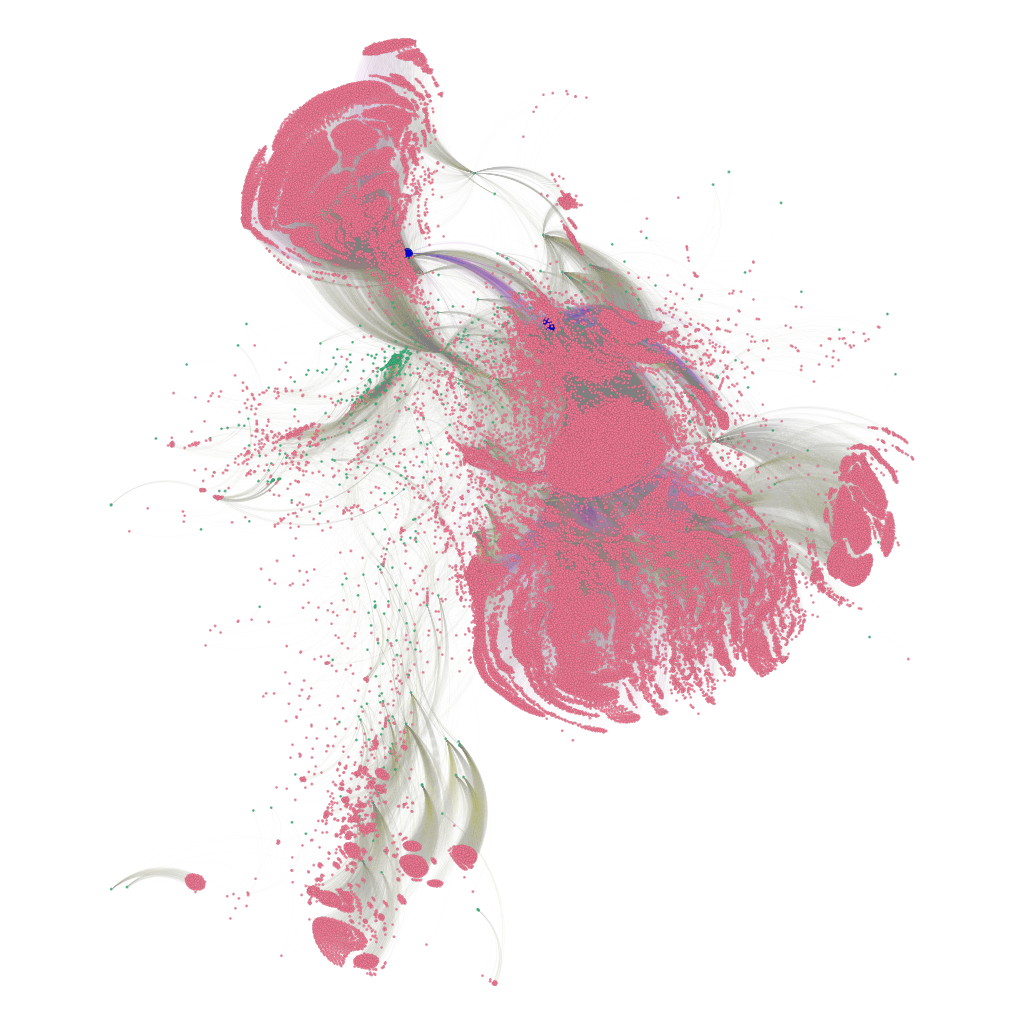}
    \caption{Here we show three example financial statements networks: on the left is a small financial statements network, the middle is an average financial statements network, and on the right is a large financial statements network. The blue nodes and edges have the highest centrality values, revealing interesting parts of the financial structure.}
    \label{fig:examplenetworks}
\end{figure}

\textcolor{black}{All the journal entry datasets are transformed into networks. We follow the method of \cite{Boersma2018} to convert the transaction data into a network. In Figure~\ref{fig:examplenetworks}, we show three selected financial statements networks ranging from small to large.}

\begin{table}[]
\centering
\begin{tabular}{@{}lll@{}}
\toprule
Industry & Description                & Number of companies \\ \midrule
CRS      & General manufacturing      & 39                  \\
HLP      & Healthcare                 & 30                  \\
LE       & Energy (incl. oil \& gass) & 19                  \\
PF       & Publishing         & 19                  \\
RTL      & Retail                     & 29                  \\ \bottomrule
\end{tabular}
\caption{For a subset of the total dataset we show the distribution of the number of companies in each industry. }
\label{table:industrycoverage}
\end{table}

\section{Network statistics}
\label{sec:basicnetworkstats}
We start with the basic network properties. First, we discuss the size of the network. Secondly, the betweenness centrality, closeness centrality, and degree centrality. 

\subsection{Network size}
The size of the network is determined by two elements: the number of nodes and the connections between the nodes. The first point is trivial; more nodes imply a larger network. To illustrate the impact of the \textcolor{black}{connectivity} on the network size, assume two networks (i) a network with 10 nodes connected as a chain, and (ii) a network with 10 nodes connected as a star. Despite that both networks have the same number of nodes, we consider network (i) to be larger than network (ii) because the longest shortest path in the network is larger -- this is the network's diameter. In this section, we investigate the size of the financial statements network by counting the number of nodes and measuring the diameter.

In the financial statements network, we have two types of nodes: financial account nodes and business process nodes. We first analyze the financial account nodes and then analyze the business process nodes. For each node type, we briefly explain the impact of the number of nodes on the financial structure -- what does the number of nodes represent in a company.

\begin{figure}
    \centering
    \includegraphics[width=0.4\textwidth]{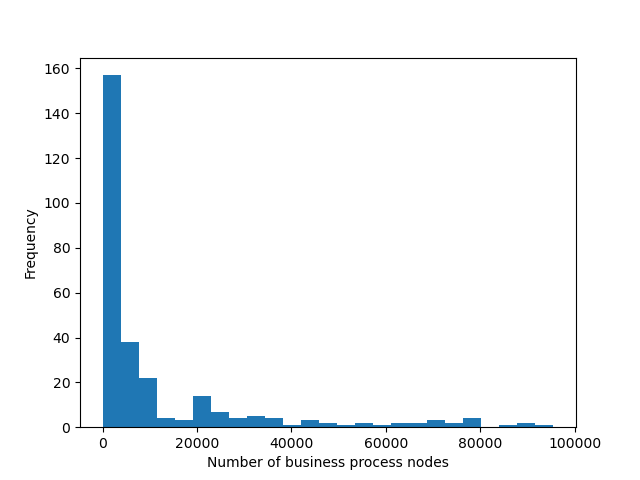}
    \includegraphics[width=0.4\textwidth]{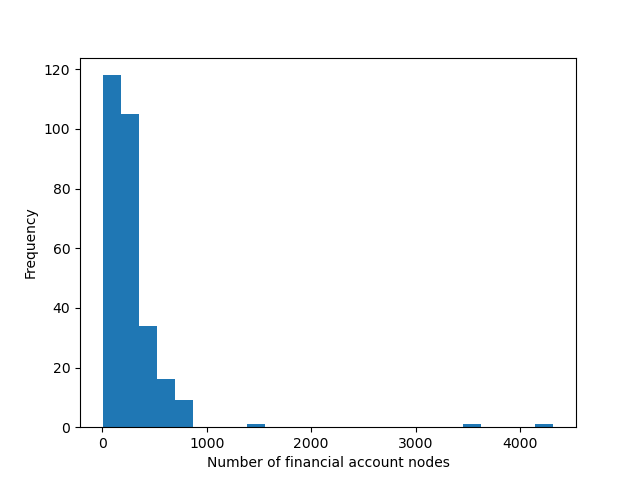}
    \caption{Left) The number of business process nodes in a financial statements network. On average, we have 11.668 nodes, the minimum is 1, and the maximum is 95.954. Right) The number of financial account nodes in a financial statements network. On average, we have 272 nodes, the minimum is 3, and the maximum is 4319.}
    \label{fig:nodes:bp}
\end{figure}

A company creates financial account nodes to report its financial position. The number of nodes can vary between companies, depending on their specific reporting needs. For example, a company may be required to oversee numerous objects, such as holiday residences within a vacation resort, and large projects with more elaborate reporting requirements, such as work-in-progress accounting. Consequently, additional financial accounts are used to track the financial position. Therefore, the number of financial account nodes provides valuable insights into the intricacy of the financial structure, which may have potential ramifications on the audit planning, given that each financial account might require a different audit approach due to its distinct nature. For the financial account nodes, we have, on average, 272 nodes. Figure~\ref{fig:nodes:bp} (right) shows the histogram of the number of financial account nodes. We observe values as low as 3 financial accounts and as high as 4.319 financial accounts. Interestingly, this suggests that 272 financial account nodes are expressive enough to measure a company's financial performance. The minimum value of 3 is most likely a partial dataset. We analyzed the networks with many financial accounts and found, for example, a more detailed revenue administration -- they tracked the revenue of many objects.

The interpretation of the number of business process nodes is more complex due to their definition. A new business process node is created when we observe a distinct pattern transaction. A distinct pattern is defined as a combination of financial account nodes~\cite{Boersma2018}. For example, a sale results in a flow between the revenue and the trade receivables financial accounts. Consequently, we add a business process node to the network for this observed pattern. But if we have another sales transaction with the same pattern, we do not add a new business process node -- the pattern is not unique. To illustrate how unique patterns emerge, consider a company with a sales and purchase department. The business activity of each department results in unique patterns of monetary flows in the network. Thus, the number of distinct patterns -- the number of business process nodes -- provides us with insights into the complexity of the financial structure. In the 285 datasets we analyzed, we observe, on average, 11.668 business process nodes. Figure~\ref{fig:nodes:bp} (left) shows the histogram of the number of business process nodes. Note that the variance is high. We have values as low as 1 business process node, but we also observe values as high as 95.954 business process nodes. The histogram shows a long tail suggesting that some companies have a high complexity compared to the average complexity observed.

\begin{figure}
    \centering
    \includegraphics[width=0.7\textwidth]{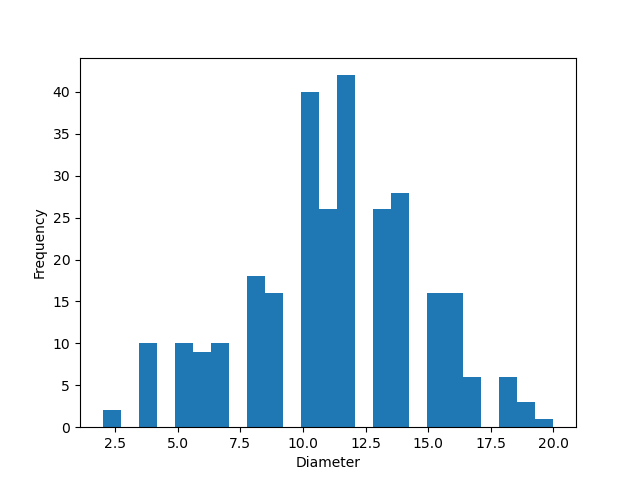}
    \caption{This is the diameter distribution of the financial statements networks. The mean diameter is 11.25 and the minimum is 2 and the maximum is 20.}
    \label{fig:diameter}
\end{figure}

Recall the example of the chain and the star connection pattern, the same number of nodes yet a different diameter. In the case of a financial statements network, the diameter provides us with insights about what kind of complexity is increasing. As mentioned before, for business process nodes, new nodes are added due to completely new distinct patterns but also due to small variations in existing patterns. For example, 90\% of the pattern is similar; however, the 10\% that is different results in a unique pattern -- think of a sales transaction with different discount rates. We refer to this as the increased complexity of the same business process. We expect that the first type of complexity yields a larger diameter while the latter does not increase the diameter -- it simply adds pathways between the same set of nodes. In Figure~\ref{fig:diameter}, we observe that, on average, the diameter is approximately 11. Interestingly,
the smallest diameter observed is 2, and the largest is 20. The diameter of 2 is likely a financial system with a single business process connecting all financial accounts. Consequently, all paths have a maximum length of 2. 

In summary, the main driver of the size of the network is the number of business process nodes, whereas the number of financial account nodes is relatively small. Surprisingly, by measuring the diameter, we learn that the maximum diameter (20) is approximately only twice the average (11.25), while the maximum number of business process nodes is approximately eight times as high (95.954) as the mean (11.668). This suggests that the financial structure organizes itself such that the longest shortest path (diameter) stays small with respect to the increasing number of nodes in the network. This suggests that paths are not extended, but more alternative paths are added between the same set of financial account nodes in the network.

\subsection{Centrality measures}
\label{sec:centrality}
Centrality measures highlight important parts of the financial statements networks, helping the auditor to pinpoint financial gateways, financial hubs, and financial core activities of a company. This section discusses the centrality measures applied to a financial statements network.

\begin{figure}
    \centering
    \includegraphics[width = 0.85\textwidth]{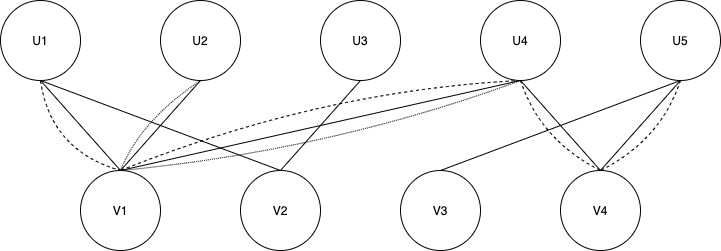}
    \caption{A bipartite network $G = (V, U, E)$ with two shortest paths between U1-U5 (dashed line) and U2-U4 (dotted line) and the betweenness of $V1$ which is part of these shortest paths. Note that the betweenness of $V1$ is only part of the two shortest paths. This is the unnormalized betweenness of $V1$. }
    \label{fig:betweenness}
\end{figure}

First, we discuss the financial gateways indicated by the betweenness centrality measure. To explain what betweenness is, assume we have all the shortest paths between all pairs of nodes. Then, for a single node, we count how often it is on the shortest path between any other pair of nodes~\cite{Freeman1977-rm}. 
In complex systems, the node with the highest betweenness often plays a key role. High betweenness implies that a node is likely a bridge between two or more communities. These gatekeepers between two or more communities are interesting to study~\cite{Borgatti1997}. 
For example, in real-world network datasets, 
\cite{Faust1997-wu} studies the betweenness measure for the actor network where a high betweenness implies that an actor can mediate resources or information between other actors.
In bipartite graphs, however, the betweenness measure will (almost) always be lower than the maximum due to the bipartite nature of the graph. 
\cite{Borgatti1997} proposes to normalize this measure in a way that respects the bipartite structure. 
Figure~\ref{fig:betweenness} shows the betweenness of a network.
The centrality betweenness is defined as~\cite{Borgatti2009-hr}:
\begin{equation}
    C_b(v_i) = \frac{1}{2}\sum_{k \neq i} \sum_{j\neq k, i}\frac{I[v_i \in d(v_k, v_j)]}{d(v_k, v_j)} 
\end{equation}
with $d(., .)$ as the shortest path, and $I[v_i \in d(v_k, v_j)]$ is an indicator function that is equal to 1 if a path passes through node $v_i$ and 0 otherwise. Figure~\ref{fig:betweenness} shows an example of a bipartite network and the betweenness statistic.
For a bipartite network $G=(V, U, E)$ with $U$ and $V$ as sets of nodes and $E$ as the set of edges, we use the following normalization~\cite{Borgatti2009-hr}:
\begin{equation}
\begin{split}
    b_{U} = \frac{1}{2}[m^2(s+1)^2 + m(s+1)(2t - s - 1) - t(2s-t+3)] \\
    s = \frac{n-1}{m} \\
    t = (n - 1) \mod m
\end{split}
\end{equation}
with $n = |U|$ and $m = |V|$.
For the other partition, we use
\begin{equation}
\begin{split}
    b_{V} = \frac{1}{2}[n^2(p+1)^2 + n(p+1)(2r - p - 1) - r(2p-r+3)] \\
    p = \frac{(m-1)}{n} \\
    r = (m - 1) \mod n
\end{split}
\end{equation}
Then we calculate it as:
\begin{equation}
C_b^*(v_i) = \begin{cases} 
      \frac{C_b(v_i)}{b_{U}}, & \text{if } v_i \in V\\
      \frac{C_b(v_i)}{b_{V}}, & \text{if }v_i \in U 
 \end{cases}
\end{equation}
This enables us to normalize the measure by the theoretical maximum and avoid the general definition of the betweenness for one-mode projections. 
We selected three networks (displayed in Figure~\ref{fig:examplenetworks}) and discuss the nodes with the highest betweenness in each network  - the financial gatekeepers. In all networks, the highest betweenness nodes are financial account nodes. More specifically, we selected two nodes that are likely trade liability accounts in the small network, a holding company. The third node relates to a foreign exchange revaluation account. Given that this is a holding company with many activities in foreign countries, it makes sense that these accounts are important to study in more detail. 
In the medium network, the nodes with the highest betweenness relate to automatic payments, creditor payments, and expense declarations. These components explain the core activity of the company because the company is a healthcare provider. The monetary flows relate to the declaration of provided care and the payments thereof. In the large network, which is a hospital, the selected nodes are purchase orders of various goods, food expenses, and payment of invoices. Despite that the core activity of a hospital is to provide care, patients also have to eat, and goods are purchased to provide the care. Interestingly, the results suggest that the majority of the financial structure's complexity is determined by the processes related to these activities. 

Secondly, we discuss the financial hubs indicated by the closeness centrality measure. To explain what closeness is, consider all the shortest paths of a single node to all other nodes in the network. The sum of these distances $f$ is the farness of that node, and the closeness is its inverse $1/f$~\cite{Freeman1978-ce}.
The closeness describes how well connected a node is to all other nodes in the network. In real-world networks, for example, the actor-network in \cite{Faust1997-wu}, they measure the closeness centrality of an actor. He argues that central actors can efficiently contact others.
In the case of a bipartite network, the minimal distance between nodes of the same set is always larger than 1. Consequently, the closeness is lower as an artifact. \cite{Borgatti1997} proposes to normalize in a way that respects the bipartite nature of the network.
For a one-mode network, the closeness centrality is defined as~\cite{Borgatti2009-hr}:
\begin{equation}
    C_c(v_i) = \frac{n-1}{\sum_y d(v_i,v_y)}
\end{equation}
with $n$ as the number of nodes and $d(.,.)$ as the shortest-path. We normalize by $n-1$, the maximum number of connections it can have.
However, for bipartite graphs, this does not work. Therefore, we normalize by the theoretical minimum, which is given by~\cite{Borgatti2009-hr}:
\begin{equation}
C_b^*(v_i) = \begin{cases} 
      \frac{|V| + 2(|U|-1)}{\sum_y d(v_i,v_y)}, & \text{if } v_i \in U\\
      \frac{|U| + 2(|V|-1)}{\sum_y d(v_j,v_y)}, & \text{if } v_i \in V 
 \end{cases}
\end{equation}
where $|U|$ is the number of nodes in $U$ and $|V|$ is the number of nodes in $V$ and $d(.,.)$ as the shortest-path.
We follow a similar approach to our analysis of betweenness centrality. We use the three networks (displayed in Figure~\ref{fig:examplenetworks}) and discuss the nodes with the highest closeness in each network -- the financial hubs. Interestingly, here we observe that the closeness measure selects both financial account nodes and business process nodes. Whenever it selects a financial account node, it is the same node as indicated by the betweenness centrality measure. In contrast to the betweenness measure, for the small, medium, and large network, it selects one, none, and two business processes, respectively.

\begin{figure}
    \centering
    \includegraphics[width = 0.45\textwidth]{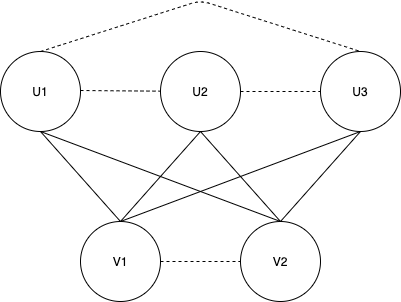}
    \caption{A bipartite network where the solid lines are the maximum number of possible edges, while in a one-mode network the dashed edges are also possible.}
    \label{fig:degree}
\end{figure}

Finally, we discuss the financial core activities as indicated by the degree centrality. The average degree and the density of a network are often reported statistics. In real-world networks, for example, \cite{DeMasi} studies the bank credit network and reports the average degree and shows that the degree distribution suggests that firms prefer lending from a single counterparty while others prefer to lend from multiple parties.
However, we must apply these measures with care for a bipartite network because otherwise, we obtain an artificially low density. The low density is caused by normalizing with a fully connected network; this includes connections between nodes of the same set, which is, by definition, not possible in a bipartite network. Figure~\ref{fig:degree} demonstrates this with an example. Here we show all possible edges when we consider the network a one-mode network and all possible connections when we consider this a bipartite network. Notice that in the one-mode network, many more edges are possible.
The degree centrality in a one-mode network is defined as~\cite{Borgatti2009-hr}:
\begin{equation}
    C_d(v_i) = \frac{k_i}{n-1}
\end{equation}
with $n$ the number of nodes. We normalize by $n-1$, the maximum number of edges a node could have when it is connected to all other nodes in the network.
We use a different normalization to calculate the degree centrality for a bipartite network. Assume that the bipartite network consists of sets $U$ and $V$, then the degree is defined as~\cite{Borgatti2009-hr}:
\begin{equation}
\begin{split}
       C_d(v_i) = \frac{k_i}{|V|} \quad \text{if } v_i \in U \\
        C_d(v_i) = \frac{k_i}{|U|} \quad \text{if } v_i \in V \\
\end{split}
\end{equation}
Thus, we normalize with the maximum number of connections to the other partition avoiding the artificially low degree centralities.
Similar to the closeness and betweenness measure, we select the nodes with the highest degree centrality from the three example networks -- the financial core activities. Here, again, we observe that most of the selected nodes are similar to the ones selected by the closeness and betweenness measure. It selects one other financial account only for the large network compared to betweenness selection.

\section{Degree distribution and scale-free property}
\label{sec:degreedistribution}
The degree distribution is an extensively studied statistic of complex networks. The degree distribution could reveal that connections between nodes do not emerge randomly. It could indicate that there are other underlying mechanisms that result in the observed connectivity in a network. For example, the preferential attachment mechanism~\cite{barabasi2012publishing}. A large body of evidence suggests that many complex networks have power-law degree distributions~\cite{clauset2009power}. This suggests that all complex systems might share a general mechanism. 

To study whether the financial statements network has a heavy-tail degree distribution. We show the degree distribution of the financial account nodes and the business process nodes. First, we show the degree distribution of three randomly selected financial statements networks and study them in detail. Next, we propose to test all financial statements networks and study whether they all have a heavy-tail degree distribution. We use the Powerlaw python package~\cite{alstott2014powerlaw, clauset2009power} to fit the degree distributions. 

\begin{figure}
    \centering
    \includegraphics[width=0.99\textwidth]{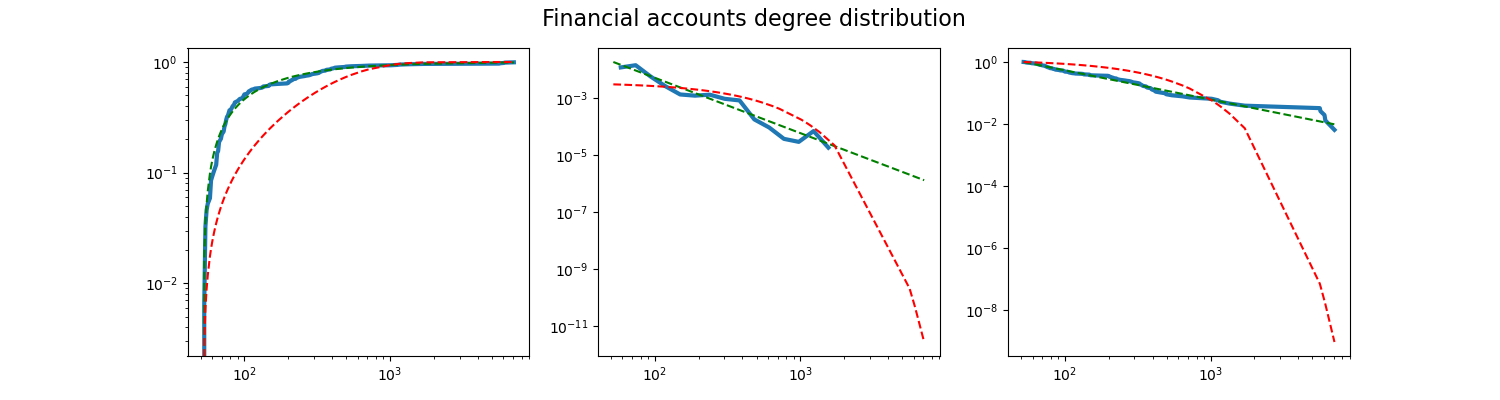}
    \includegraphics[width=0.99\textwidth]{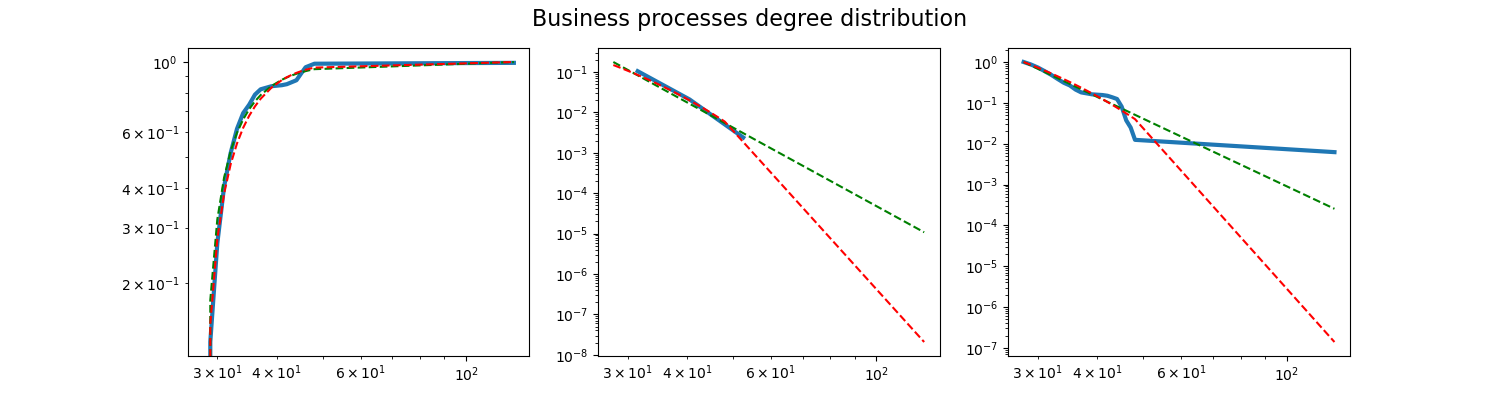}
    \caption{On the left is the cumulative density function, in the middle is the probability density function, and on the right is the complementary cumulative density function for financial statements network 250 (small) with 13.538 nodes in total. The top row is the financial account nodes distributions, and the bottom row is the business process nodes distributions. The green dashed line is a power-law distribution, the red dashed line is an exponential distribution, and the solid blue line is the empirical distribution. Note that the axes are on a log scale.}
    \label{fig:degreedist:small}
\end{figure}

\begin{figure}
    \centering
    \includegraphics[width=0.99\textwidth]{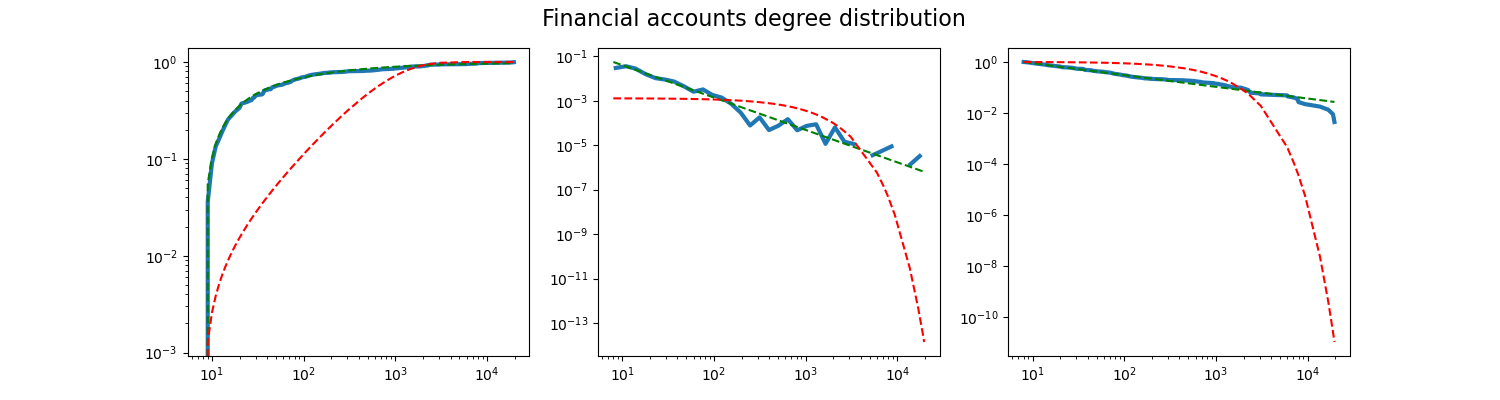}
    \includegraphics[width=0.99\textwidth]{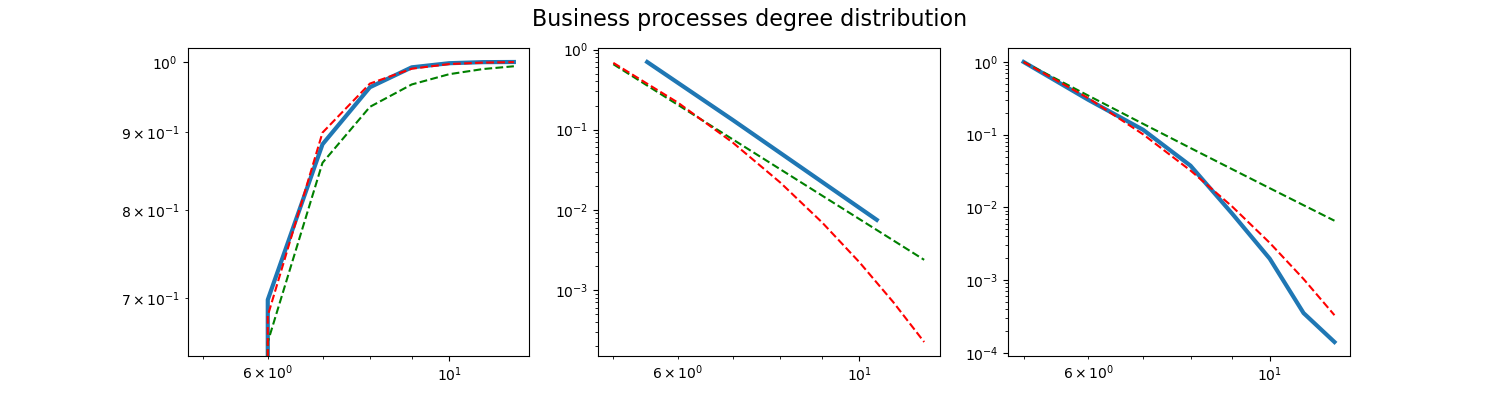}
    \caption{On the left is the cumulative density function, in the middle is the probability density function, and on the right is the complementary cumulative density function for financial statements network 322 (medium) with 45.124 nodes in total. The top row is the financial account nodes distributions, and the bottom row is the business process nodes distributions. The green dashed line is a power-law distribution, the red dashed line is an exponential distribution, and the solid blue line is the empirical distribution. Note that the axes are on a log scale.}
    \label{fig:degreedist:medium}
\end{figure}

\begin{figure}
    \centering
    \includegraphics[width=0.99\textwidth]{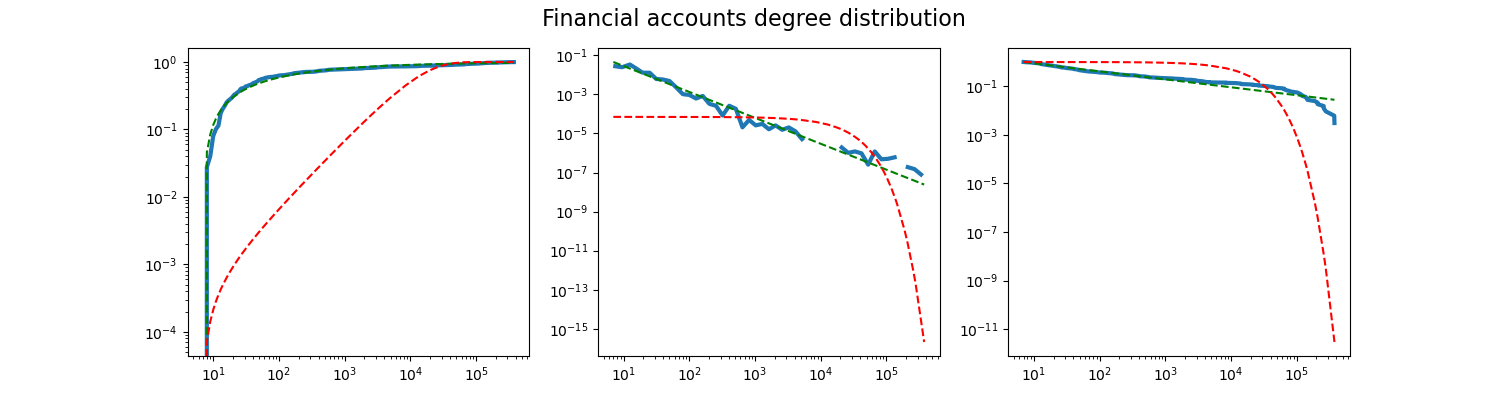}
    \includegraphics[width=0.99\textwidth]{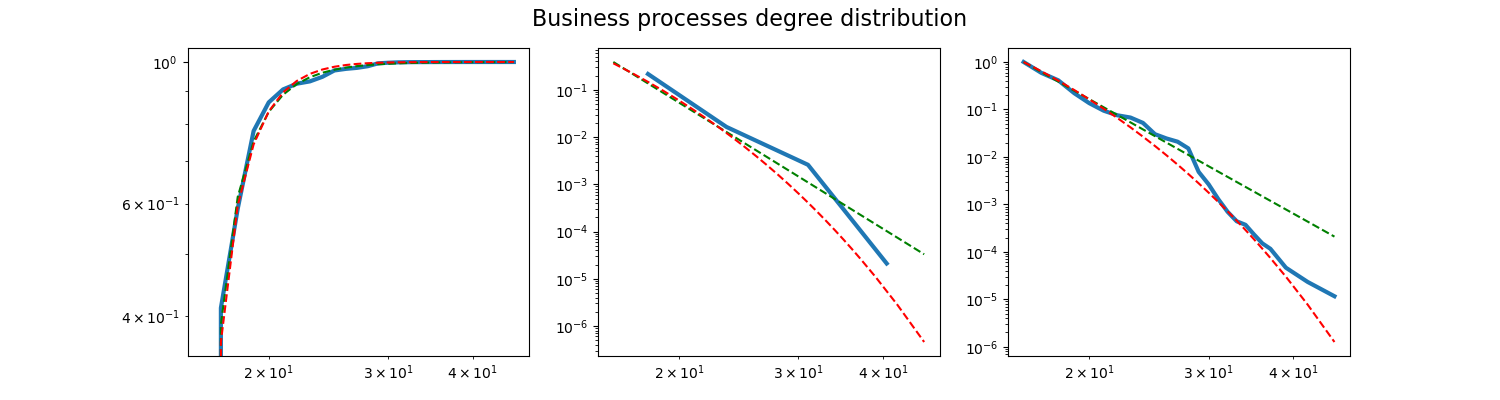}
    \caption{On the left is the cumulative density function, in the middle is the probability density function, and on the right is the complementary cumulative density function for financial statements network 62 (large) with 391.688 nodes in total. The top row is the financial account nodes distributions, and the bottom row is the business process nodes distributions. The green dashed line is a power-law distribution, the red dashed line is an exponential distribution, and the solid blue line is the empirical distribution. Note that the axes are on a log scale.}
    \label{fig:degree:dist:large}
\end{figure}

In Figure~\ref{fig:degreedist:small}, ~\ref{fig:degreedist:medium}, and ~\ref{fig:degree:dist:large}, we show the fitted degree distributions for a small, medium, and large financial statements network. For each network, we show the fit of the probability density function, the cumulative density function, and the complementary cumulative density function for an exponential and power-law distribution. We observe that the power law better fits the financial account nodes than an exponential distribution. For the business process nodes, the evidence is less conclusive.

To investigate all the financial statements networks, we propose to formulate a hypothesis test that measures the likelihood ratio L~\cite{clauset2009power}: 
\begin{equation}
    L = \prod^n_{i=1} \frac{p_1(x_i)}{p_2(x_i)}
\end{equation}
where $p_1(.)$ and $p_2(.)$ are distributions and $x_i$ is an observed value and $n$ is the number of observations. In the hypothesis, we assume under the null hypothesis an exponential distribution ($p_1(.)$), and for the alternative hypothesis a power-law distribution ($p_2(.)$). To compute the statistic we take the log-likelihood as discussed in \cite{clauset2009power} and \cite{alstott2014powerlaw}. Note that we can determine the p-value of this test~\cite{clauset2009power}. When the $p$-value is small the test is reliable. Thus, we can conclude with more certainty which distribution is preferred. In the case of a high p-value, the test cannot discriminate between the two distributions. If the likelihood ratio $L$ is larger than 1, then the likelihood of $p_1(.)$ is higher than $p_2(.)$. Therefore, $p_1(.)$ is preferred over $p_2(.)$. This does not provide us with a good fit in absolute terms, but merely that one distribution is a better description of the data than the other. 

We selected the exponential distribution as the null hypothesis and not the power law because, in case of a rejection, we reject the exponential distribution in favor of a power-law distribution. Note that the other way around does not hold. When we fail to reject the null hypothesis, we cannot conclude that an exponential distribution is a good fit. Thus, we can find supporting evidence of a heavy-tailed distribution with this experimental setup. 

Table~\ref{tab:powerlaw} shows the percentage of tests (with a p-value lower than 0.1) that prefer the power-law distribution over the exponential distribution. This suggests that the power law is often preferred for both the financial account nodes' degree distribution as well as the business process nodes' degree distribution. However, the evidence for the financial account nodes' degree distribution is stronger. Interestingly, Table~\ref{tab:powerlaw:industry} shows the same information but is categorized per industry type. Note that the HLP industry shows strong evidence that both node types show a power-law degree distribution. The RTL industry, in contrast, prefers the exponential distribution for the business process nodes. This suggests that different industries have different degree distributions for the business process nodes. For the financial account nodes, they are all in agreement that the distribution is heavy-tailed.

\begin{table}[]
    \centering
    \begin{tabular}{@{}ll@{}}
    \toprule
              & Power-law vs. exponential \\ \midrule
    Degree FA & 85\%                      \\
    Degree BP & 68\%                      \\ \bottomrule
    \end{tabular}
    \caption{Hypothesis test results, the percentage of tests that reject the null hypothesis (exponential distribution) in favor of the alternative (power-law distribution).}
    \label{tab:powerlaw}
\end{table}

\begin{table}[]
    \centering
    \begin{tabular}{@{}lll@{}}
    \toprule
    Industry & FA power vs. exp & BP power vs. exp \\ \midrule
    CRS      & 94\%             & 67\%             \\
    HLP      & 100\%            & 95\%             \\
    RTL      & 100\%            & 26\%             \\
    LE       & 100\%            & 50\%             \\
    PF       & 100\%            & 63\%             \\ \bottomrule
    \end{tabular}
    \caption{Hypothesis test results categorized per industry type, the percentage of tests that reject the null hypothesis (exponential distribution) in favor of the alternative (power-law distribution).}
    \label{tab:powerlaw:industry}
\end{table}

\section{Discussion and conclusion}
\label{sec:conclusion}
We collected 300+ datasets with transaction data of companies. The transaction data is converted into  financial statements networks which we analyze to obtain a better understanding of the financial structure of companies.

Our study reveals that the number of business process nodes determines a financial structure's size, and an increase in the number of nodes does not result in a corresponding proportional increase in the network's diameter. Moreover, we found evidence of a heavy-tailed degree distribution in financial account nodes, leading to hubs of interest to auditors. Results confirm, for a small sample, that the centrality measures highlight important hubs in the financial structure, increasing the auditor's understanding of the company.

Thus, the characteristics of the baseline network statistics increased our understanding of a company's financial structure. In the future, these baseline statistics could enable auditors to detect anomalies in their structural properties. 
For future work, instead of calculating the network measures for each node, it would be interesting whether we can derive useful node embeddings. Finally, we only found weak evidence to support the claim that industries have unique characteristics. This, however, might be an artifact of the measure selected. Therefore, it would be interesting to apply other measures such as the Weisfeiler-Lehman~\cite{Yu_Weisfeiler1968-lt} propagation that is used to detect graph isomorphism -- a potential way to detect similarities across networks~\cite{togninalli2019wasserstein}.

\par

\bibliography{references_manual.bib}

\section*{Author contributions statement}
M.Boersma wrote the manuscript. M.Boersma conducted the experiment(s), all authors (M.Boersma, S.Sourabh, L.A. Hoogduin, D. Kandhai) analysed the results.  All authors (M.Boersma, S.Sourabh, L.A. Hoogduin, D. Kandhai) reviewed the manuscript.

\section*{Declaration of interest and disclaimer:} The authors report no conflicts of interest, and declare that they have no relevant or material financial interests related to the research in this paper. The authors alone are responsible for the content and writing of the paper, and the views expressed here are their personal views and do not necessarily reflect the position of their employer.

\end{document}